\documentclass[aps,pre,twocolumn,superscriptaddress,showpacs,float]{revtex4-1}
\usepackage[utf8]{inputenc}
\usepackage{amsmath}
\usepackage{amssymb}
\usepackage{latexsym}
\usepackage{ifpdf}
\usepackage{esint}
\usepackage{dsfont}
\usepackage{slashed}
\usepackage{color}

\ifpdf
\usepackage[pdftex]{graphicx}
\else
\usepackage{graphicx}
\fi

\ifpdf
\DeclareGraphicsExtensions{.pdf, .jpg, .tif}
\else
\DeclareGraphicsExtensions{.eps, .jpg}
\fi

\def\imag{{i}}

\renewcommand{\Re}{\mathrm{Re}}
\renewcommand{\Im}{\mathrm{Im}}

\begin{document}

\title{Derivation of the Lifshitz-Matsubara sum formula
for the Casimir pressure between metallic plane mirrors}

\author{R. Gu\'erout} \email[]{romain.guerout@upmc.fr}
\affiliation{Laboratoire Kastler-Brossel, CNRS, ENS, UPMC, Case 74,
F-75252 Paris, France}
\author{A. Lambrecht}
\affiliation{Laboratoire Kastler-Brossel, CNRS, ENS,
UPMC, Case 74, F-75252 Paris, France}
\author{K. A. Milton}
\affiliation{Laboratoire Kastler-Brossel, CNRS, ENS, UPMC, Case 74,
F-75252 Paris, France} \affiliation{H. L. Dodge Dept. of Physics and
Astronomy, Univ. of Oklahoma, Norman, OK 73019 USA}
\author{S. Reynaud}
\affiliation{Laboratoire Kastler-Brossel, CNRS, ENS, UPMC, Case 74,
F-75252 Paris, France}
\date{\today}

\begin{abstract}
We carefully re-examine the conditions of validity for the
consistent derivation of the Lifshitz-Matsubara sum formula for the
Casimir pressure between metallic plane mirrors. We recover the
usual expression for the lossy Drude model, but not for the lossless
plasma model. We give an interpretation of this new result in terms
of the modes associated with the Foucault currents which play a role
in the limit of vanishing losses, in contrast to common
expectations.
\end{abstract}


\maketitle

\section{Introduction}
\label{sec:introduction}

The Casimir force~\cite{Casimir1948} is a manifestation of vacuum
field fluctuations~\cite{Milton2001} which is now measured with a
good experimental precision in various
experiments~\cite{Klimchitskaya2009,Lamoreaux2011,Capasso2011,Decca2011}.
However, the comparison of experimental results with theoretical
predictions remains a matter of
debate~\cite{Klimchitskaya2006,Brevik2006,Lambrecht2011}. The
original Casimir formula had a universal form, with the pressure
between two plane plates being $P=-\hbar c\pi^2/240 L^4$ as a
function of the inter-plate distance $L$, because the mirrors were
idealized as perfectly reflecting and thermal fluctuations were
ignored. But experiments are performed with imperfect reflectors, at
room temperature, so that the experimental results have to be
compared with the Lifshitz
formulas~\cite{Lifshitz1956,Dzyaloshinskii1961} which take these
effects into account.

Most experiments are performed with mirrors covered by thick layers
of gold, and their optical properties are described by reflection
amplitudes calculated from Fresnel equations at the interfaces
between vacuum and metallic bulks~\cite{Jaekel1991}. These
reflection amplitudes are deduced from a frequency-dependent
dielectric function $\varepsilon(\omega)$, which is the sum of
contributions corresponding to bound electrons and conduction
electrons. The function $\varepsilon(\omega)$ is deduced from
tabulated optical data~\cite{Lambrecht2000,Svetovoy2008} and
extrapolated to low frequencies by using the Drude model for
describing the conductivity of gold, $\sigma (\omega) =
\omega_\mathrm{p}^2/(\gamma-\imag \omega)$, where
$\omega_\mathrm{p}$ is the plasma frequency and $\gamma$ the damping
parameter. This model incorporates the important fact that gold has
a finite static conductivity $\sigma_0 =
\omega_\mathrm{p}^2/\gamma$.

The limiting case of a lossless plasma of conduction electrons
($\gamma = 0$) is also often considered. This model cannot be an
accurate description of metallic mirrors as it contradicts the fact
that gold has a finite static conductivity while leading to a poor
extrapolation of tabulated optical data. However, as $\gamma$ is
much smaller than $\omega_\mathrm{p}$ for a good metal such as gold
and the effect of dissipation is appreciable only at low frequencies
$\omega\lesssim\gamma$ where $\varepsilon$ is very large for both
models, one might expect that dissipation does not affect
significantly the value of the Casimir force. This naive expectation
is met at small distances or low temperatures but not in the general
case. In fact, dissipation has a significant effect on the value of
the Casimir force at room temperature at distances accessible in
experiments~\cite{Bostrom2000,Ingold2009,Brevik2014}. Furthermore,
some experimental results appear to lie closer to the predictions of
the lossless plasma model than to that of the dissipative Drude
model~\cite{Decca2007prd,Decca2007epj,Chang2012}. Other experiments
at larger distances, $L>1\mu$m, have led to a better agreement with
the dissipative model~\cite{Sushkov2011,Milton2011}, at the price of
a large correction due to the effect of electrostatic
patches~\cite{Kim2010}. This weird status of theory-experiment
comparison has led to a large number of contributions, and many
references can be found in the lecture notes~\cite{Reynaud2014}.
Among a variety of ideas, it has been suggested that the Lifshitz
formulas might not be valid for dissipative media~\cite{Bordag2011}.

The aim of the present paper is to check carefully the conditions of
validity for the whole derivation of the Lifshitz formulas for the
Casimir pressure between metallic plane mirrors, in particular for
the two cases of the lossy Drude model and lossless plasma model. We
focus attention on the questions related to the discontinuities
appearing at the limit $\gamma\to0$ of vanishing dissipation. In
particular, we discuss with great care the equivalence of two kinds
of Lifshitz formulas. The first one, which we will call the Lifshitz
formula in the following, is an integral over all field modes
characterized by real frequencies, while the second one, which we
will call the Lifshitz-Matsubara formula, is a discrete sum over
purely imaginary Matsubara frequencies~\cite{Matsubara1955}.

We focus the discussion on the case of plane mirrors made of
non-magnetic matter. We do not treat the problems associated with
experiments performed in the plane-sphere geometry and also
disregard the discussion of possible systematic effects in the
theory-experiment comparison. References can be found
in~\cite{Lambrecht2011,Reynaud2014} for general discussions,
in~\cite{Banishev2012,Banishev2013} for experiments with magnetic
mirrors, in~\cite{Behunin2012} and~\cite{Broer2012} for systematic
effects due to electrostatic patches and roughness respectively.


\section{The Casimir radiation pressure between plane mirrors}
\label{sec:casimirpressure}

We consider two plane and parallel mirrors placed in electromagnetic
vacuum and forming a Fabry-Perot cavity. All fields in the outer or
inner regions of this cavity can be deduced from the reflection
amplitudes of the mirrors. The radiation pressures are different on
the inner and outer sides of the mirrors, and the Casimir force is
just the result of this difference integrated over all field
modes~\cite{Jaekel1991}. This approach is valid for lossy as well as
lossless mirrors~\cite{Genet2003,Lambrecht2006}, provided thermal
equilibrium holds for the whole system, so that all input
fluctuations, coming from electromagnetic fields, electrons, phonons
or any loss mechanism, correspond to the same temperature $T$. The
expression, to be written in the next paragraph, is valid and
regular for any optical model of mirrors obeying causality and high
frequency transparency properties. It reproduces the Lifshitz
formulas~\cite{Lifshitz1956,Dzyaloshinskii1961} when the mirrors are
described by reflection amplitudes deduced from Fresnel equations,
and also goes to the ideal Casimir expression when the mirrors tend
to perfect reflection~\cite{Schwinger1978}.

The expression obtained in this manner for the Casimir pressure $P$
is a sum over all modes, that is, an integral over the field
frequency $\omega$ and the transverse components $\mathbf{k}$ of the
wavevector and a sum over the polarizations $\varsigma$
\begin{eqnarray}
\label{eq:pressureg} && P = \sum_\mathbf{k} \sum_\varsigma
{\int}_0^\infty \frac{\mathrm{d}\omega}{2\pi} ~ \hbar k_z \left(
g_\mathbf{k}^\varsigma(\omega) - 1 \right) C(\omega) ~, \\
&& g_\mathbf{k}^\varsigma(\omega)\equiv \frac{1-\vert
\rho_\mathbf{k}^\varsigma (\omega) \vert^2} {\vert 1 -
\rho_\mathbf{k}^\varsigma (\omega) \vert^2} ~,\;
\rho_\mathbf{k}^\varsigma (\omega) \equiv
\left(r_\mathbf{k}^\varsigma(\omega)\right)^2 e^{2\imag k_z L}
~,\nonumber \\
&& C(\omega) \equiv \coth\frac{\hbar\omega}{2k_\mathrm{B} T} =
1+2n_\omega ~,\;
n_\omega=\frac1{\exp\frac{\hbar\omega}{k_\mathrm{B}T} -
1}~.\nonumber
\end{eqnarray}
The sum over $\mathbf{k}$ is in fact a double integral over the
components $(k_x,k_y)$ in the plane of the mirror (with the normal to the cavity
along the $z$-direction) $\sum_\mathbf{k}\equiv\iint\mathrm{d}k_x
\mathrm{d}k_y/(4\pi^2)$ while the sum over $\varsigma$ is on TM
(transverse magnetic) and TE (transverse electric) polarizations.
The function $C(\omega)$ represents the equivalent number of photons
per mode corresponding to vacuum and thermal fluctuations which
impinge the cavity from its two sides (with $n_\omega$ the number of
thermal photons in Planck's law). The function
$g_\mathbf{k}^\varsigma(\omega)$ is the ratio of the energy density
inside the cavity to that outside for a given mode. It is deduced
from the reflection amplitudes $r_\mathbf{k}^\varsigma(\omega)$ of
the two mirrors, supposed to be identical for the sake of
simplicity, and the propagation factor $\exp(2 \imag k_z L)$, with
$k_z$ the longitudinal component of the wavevector. The integral
over frequencies includes the contributions of propagative ($\omega>
c\vert\mathbf{k}\vert$) and evanescent ($\omega <
c\vert\mathbf{k}\vert$) waves, with $k_z=
\sqrt{\omega^2/c^2-\mathbf{k}^2}$ and $k_z= \imag
\sqrt{\mathbf{k}^2-\omega^2/c^2}$ respectively. Transverse
wavevectors and polarizations are preserved in the situation
considered in the present paper, and thus remain spectators
throughout the discussions.

Resonant modes correspond to an increase of energy in the cavity
with $g_\mathbf{k}^\varsigma(\omega)>1$ and they produce repulsive
contributions to the pressure. In contrast, modes out of resonance
correspond to a decrease of energy in the cavity with
$g_\mathbf{k}^\varsigma(\omega)<1$ and produce attractive
contributions. The net pressure is the balance of all contributions
after integration over modes. It is finite for any model of mirrors
and attractive between two non-magnetic mirrors. These properties
are seen more easily by rewriting the pressure $P$ as a Matsubara
sum, which is done in the following by rewriting
\eqref{eq:pressureg} in terms of an analytic function.

To this end, we introduce the closed loop function
$f_\mathbf{k}^\varsigma(\omega)$ which is a retarded causal function
associated with the Fabry-Perot cavity simply
expressed~\cite{Jaekel1991} in terms of the open loop function
$\rho_\mathbf{k}^\varsigma(\omega)$
\begin{eqnarray}
&&g_\mathbf{k}^\varsigma(\omega) = 1 +
f_\mathbf{k}^\varsigma(\omega) +
(f_\mathbf{k}^\varsigma(\omega))^\ast = 1 +
f_\mathbf{k}^\varsigma(\omega)
+ f_\mathbf{k}^\varsigma(-\omega) ~, \nonumber \\
&&f_\mathbf{k}^\varsigma(\omega)\equiv
\frac{\rho_\mathbf{k}^\varsigma(\omega)}
{1-\rho_\mathbf{k}^\varsigma(\omega)}  ~.
\end{eqnarray}
We then use the properties of this function to deduce equivalent
expressions of the Casimir pressure
\begin{eqnarray}
\label{eq:pressure} &&P = \sum_\mathbf{k,\varsigma}
\int_0^\infty \frac{\mathrm{d}\omega}{2\pi} ~
2\Re[p_\mathbf{k}^\varsigma] =
\sum_\mathbf{k,\varsigma}
\int_{-\infty}^\infty \frac{\mathrm{d}\omega}{2\pi}
\Re[p_\mathbf{k}^\varsigma]
~, \nonumber \\
&&p _\mathbf{k}^\varsigma(\omega) \equiv \hbar k_z
f_\mathbf{k}^\varsigma (\omega) C(\omega) ~.
\end{eqnarray}
In the following, we will consider the contribution
to the Casimir pressure $P$ for given values $\varsigma,\mathbf{k}$
as the integral over the real axis of the function
$p_\mathbf{k}^\varsigma$, and we will use its analyticity
properties. To this end, we will add to the integral the
contribution from its imaginary part $\Im[p_\mathbf{k}^\varsigma]$.
As the latter shows singularities on the real axis, a proper
definition of the integral will require the use of Cauchy's
principal value as discussed further below.

In order to exploit analytic properties, we introduce the complex
variable $z=\omega+i \xi$ extending $\omega$ to the complex plane.
The function $f_\mathbf{k}^\varsigma$ is defined from causal
reflection amplitudes and propagation factors, and has its poles in
the lower half of the complex plane which correspond to resonances
of the Fabry-Perot cavity. Meanwhile the function $C$ has its poles
at the Matsubara frequencies regularly spaced on the imaginary axis
\begin{eqnarray}
z_n= \imag\xi_n \quad,\; \xi_n = n \frac{2\pi k_\mathrm{B} T} \hbar
~. \label{eq:matsubarapoles}
\end{eqnarray}
Occasionally, we will also have to take care of the branch cuts in
$p_\mathbf{k}^\varsigma$ arising from the term $k_z$.

We then transform the Lifshitz formula \eqref{eq:pressure} into a
Lifshitz-Matsubara expression by a proper application of Cauchy's
residue theorem. The new expression is a sum of the residues of
$p_\mathbf{k}^\varsigma $ at the Matsubara poles
\begin{eqnarray}
\label{eq:pressureLM} &&P=-2k_\mathrm{B}T \sum_\mathbf{k}
\sum_\varsigma \sum_n^\prime \kappa_n ~
f_\mathbf{k}^\varsigma[\imag \xi_n] ~, \\
&&\kappa_n=\sqrt{\mathbf{k}^2+\frac{\xi_n^2}{c^2}} ~.\nonumber
\end{eqnarray}
The symbol $\kappa_n$ corresponds to the continuation of $k_z$ to
the Matsubara poles on the imaginary axis while the primed sum
symbol means that the contribution of the zeroth pole $n=0$ is
counted with only one half weight
\begin{eqnarray}
\sum_n^\prime \varphi(n) \equiv \frac12 \varphi(0) +
\sum_{n=1}^\infty \varphi(n)~. \label{eq:primedsum}
\end{eqnarray}

The two formulas \eqref{eq:pressure} and \eqref{eq:pressureLM} are
commonly considered as completely equivalent expressions of a single
quantity, the Casimir pressure. In the next sections, we check
carefully the conditions of validity of this equivalence property
and prove that they are indeed met when the Drude model is used, but
not when the lossless plasma model is used. We also give an
interpretation of the difference.


\section{The Drude model}
\label{sec:drudemodel}

We come now to the discussion of mirrors described by the Drude
model. To be specific, we model the permittivity of the metallic
slab as
\begin{eqnarray}
\varepsilon(\omega)=1-\frac{\omega_\mathrm{p}^{2}}{\omega(\omega+i
\gamma)} ~. \label{eq:drudemodel}
\end{eqnarray}
We thus disregard the contribution of bound electrons which do not
play an important role in discussions focused around zero
frequency. Of course, the contribution of bound electrons is taken
into account in the comparison of experiment and
theory~\cite{Decca2007prd}.

We consider that the slabs are thick enough so that the reflection
amplitudes are given by Fresnel equations at the first interface
\begin{eqnarray}
r_\mathbf{k}^\mathrm{TE}(\omega)=\frac{k_z-K_z}{k_z+K_z} ~,\;
r_\mathbf{k}^\mathrm{TM}(\omega) =\frac{\varepsilon k_z-K_z}
{\varepsilon k_z+K_z} ~,
\end{eqnarray}
where $K_z$ and $k_z$ are the longitudinal wavevectors in matter and
vacuum respectively, that is for propagating waves
\begin{eqnarray}
K_z = \sqrt{\varepsilon\frac{\omega ^2}{c^2}-\mathbf{k}^2} ~,\; k_z
= \sqrt{\frac{\omega ^2}{c^2}-\mathbf{k}^2} ~.
\end{eqnarray}
We now enter into a more detailed discussion of the poles of the
functions $f_\mathbf{k}^\varsigma$, which are the resonances of the
Fabry-Perot cavity, and lie in the lower half of the complex plane
$\Im z < 0$.

In the TM case, there are propagating Fabry-Perot modes quantized
thanks to reflection on the mirrors, as well as modes due to
hybridization of the surface plasmons living at the interfaces
between vacuum and each metallic slab, which are coupled by the
evanescent modes between the two
slabs~\cite{Intravaia2005,Intravaia2007}. The so-called $\omega_-$
mode is always evanescent with an attractive contribution to the
pressure (contrary to any other modes whose contributions are always
repulsive), while the $\omega_+$ mode can also be considered as the
first of the set of Fabry-Perot modes, with a transition from the
propagative to the evanescent sector as a function of $\mathbf{k}$.
Finally, there exists modes associated with Foucault currents which
lie on the negative imaginary axis.

\begin{figure}[htbp]
  \begin{center}
    \includegraphics[width=3.2in]{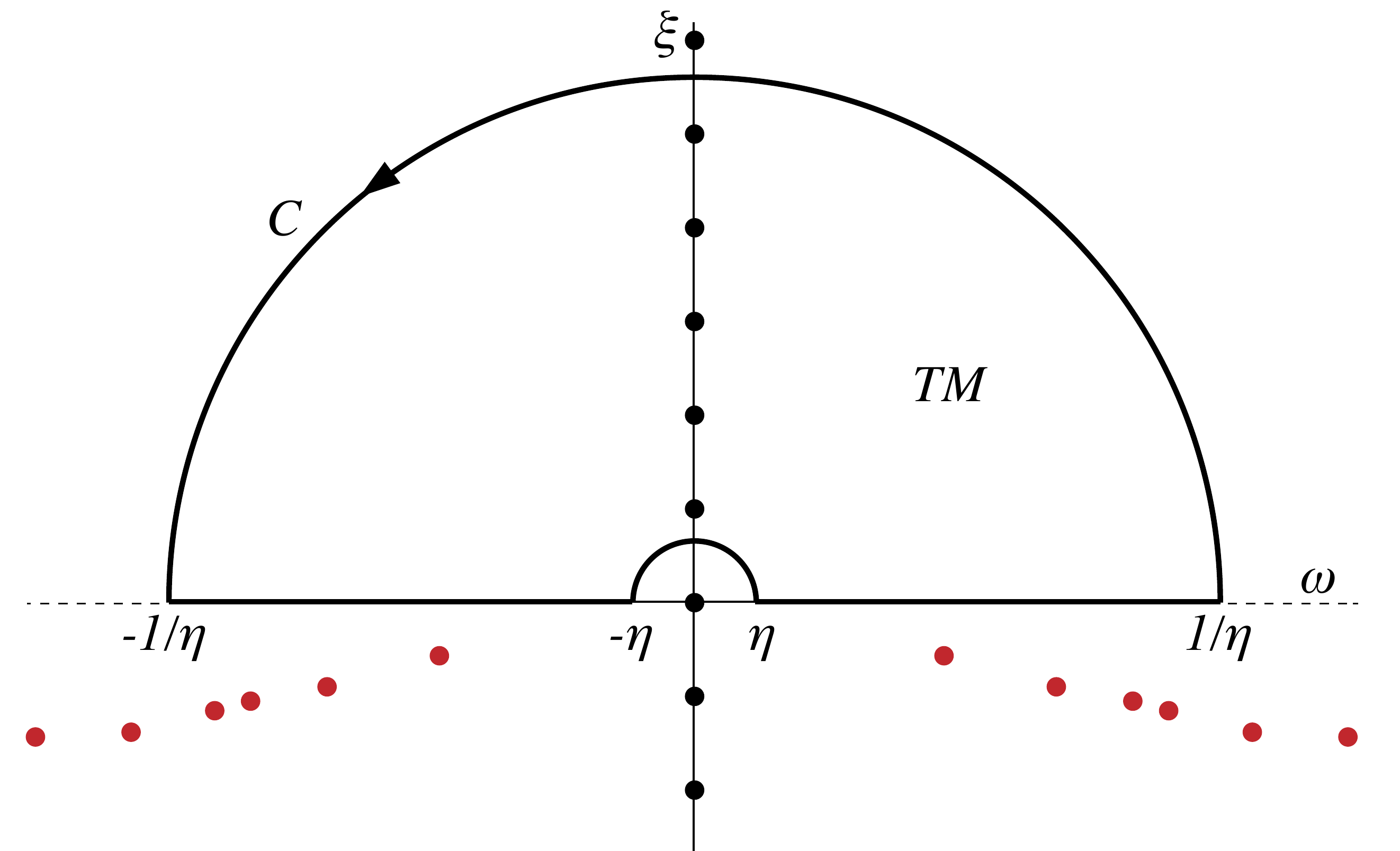}
    \includegraphics[width=3.2in]{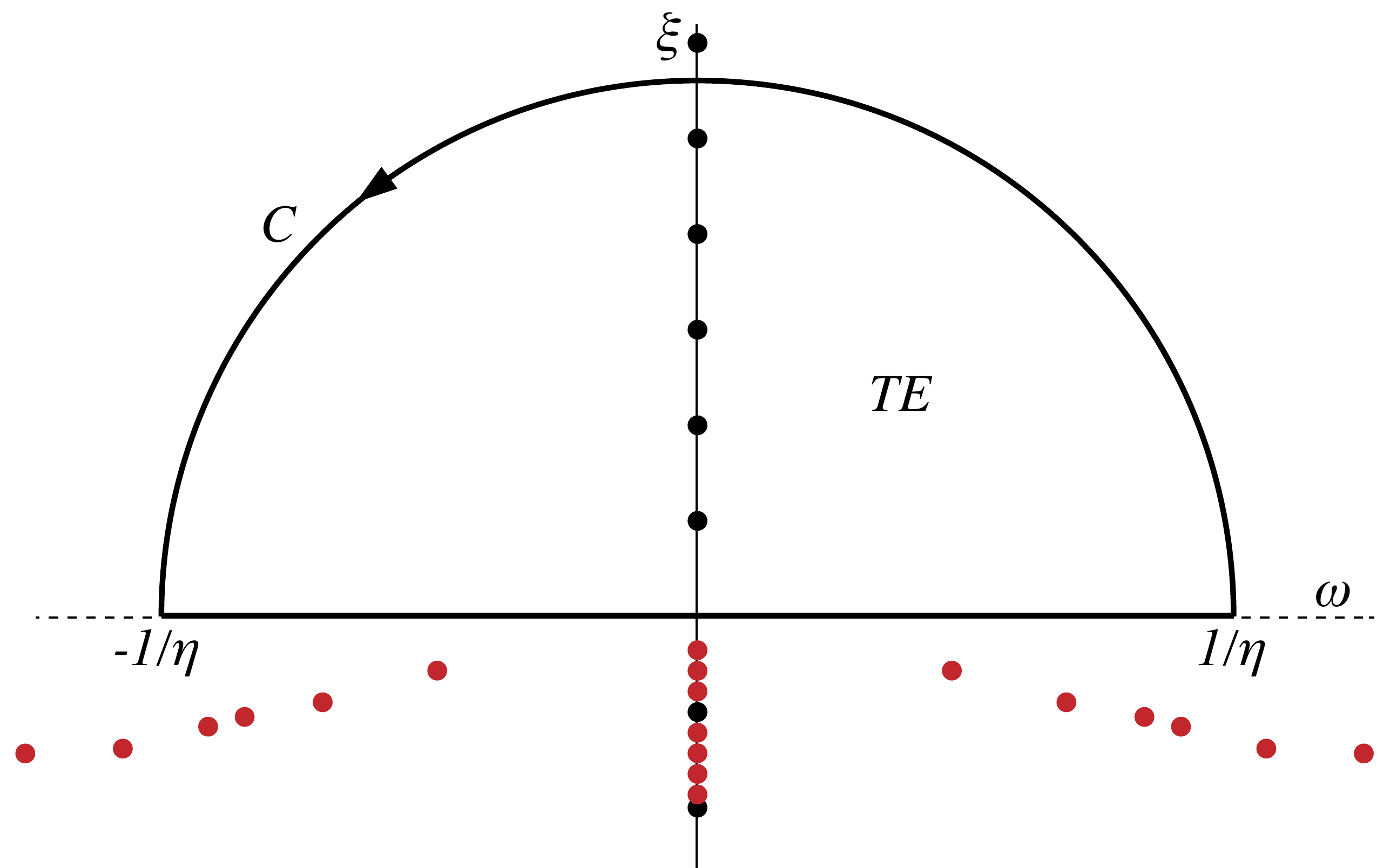}
  \end{center}
\caption{Poles of the function $f_\mathbf{k}^\mathrm{TM}(z)$ (top
plot) and $f_\mathbf{k}^\mathrm{TE}(z)$ (bottom plot) represented as
the red dots for metallic slabs described by the Drude model. Black
dots are the poles due to $C(z)$. The contours used for the
application of Cauchy's theorem are also shown. The contours pass
above the branch line associated with $k_{z}$ which runs on the real
axis for $\omega^{2}>c^{2}\mathbf{k}^{2}$. [Colors online]}
\label{fig:contourDrude}
\end{figure}

In the TE case, the situation is similar for the propagating
Fabry-Perot modes, there are no plasmonic modes but there also exist
modes arising from the interaction between the Foucault currents
living in the two slabs~\cite{Intravaia2009,Intravaia2010}. All
poles of $f_\mathbf{k}^\varsigma$ are represented as the red dots on
Fig.~\ref{fig:contourDrude} for $\varsigma=$ TM and TE on the top
and bottom plot respectively. The poles of $C$ are represented as
the black dots on the figures and the contours used below for the
application of Cauchy's residue theorem are also shown. The plots
are drawn for exaggerated values of the parameter $\gamma$, in order
to show that the red dots are below the real axis, thanks to the
finite value of $\gamma$.

The comparison of the two figures shows important differences
between the two cases.
For the TE polarization, there is no pole at the origin
$z=0$, because the behavior of $k_z f_\mathbf{k}^\mathrm{TE} \propto
z^2$ around this point leads to the disappearance of the pole in
$C(z) \propto 1/z$. We note at this point that the Foucault modes
have been shown on Fig.~\ref{fig:contourDrude} as a discrete set of
poles, which corresponds to the case of metallic slabs of finite
width $d$. This point is discussed in more detail in the next
section.

Before going further, we have to study the parity properties of the
functions involved in this discussion. We note that the permittivity
is a real function in the space-time domain, so that
$\left(\varepsilon(\omega)\right)^\ast=\varepsilon(-\omega)$. Our
choice of definition for the square roots is
such that $\left(k_z(\omega)\right)^\ast = -k_z(-\omega)$, and it
follows that $\left(p _\mathbf{k}^\varsigma (\omega)\right)^\ast=p
_\mathbf{k}^\varsigma (-\omega)$. Then, the real parts of $p
_\mathbf{k}^\varsigma$ are even functions of $\omega$ and their
imaginary parts are odd functions. After a continuation to the
complex plane, this property is read as a mirror symmetry property
with respect to the imaginary axis $\left(p _\mathbf{k}^\varsigma
(z)\right)^\ast = p _\mathbf{k}^\varsigma (-z^\ast)$, so that the
positions of poles and zeros of these functions are symmetric with
respect to the imaginary axis.

Using these parity properties as well as the properties already
discussed, we deduce
\begin{eqnarray}
\label{eq:parityDrude} &&\int_{-\infty}^{\infty} \Re[p
_\mathbf{k}^\mathrm{TE} (\omega)] \frac{\mathrm{d}\omega}{2\pi} =
2\int_{0}^{\infty} \Re[p _\mathbf{k}^\mathrm{TE} (\omega)]
\frac{\mathrm{d}\omega}{2\pi} ~, \nonumber\\
&&\int_{-\infty}^{\infty} \Im[p _\mathbf{k}^\mathrm{TE}
(\omega)] \frac{\mathrm{d}\omega}{2\pi} =0 ~, \nonumber\\
&&\int_{-\infty}^{\infty} \Re[p _\mathbf{k}^\mathrm{TM} (\omega)]
\frac{\mathrm{d}\omega}{2\pi} =2\int_{0}^{\infty} \Re[p
_\mathbf{k}^\mathrm{TM} (\omega)]
\frac{\mathrm{d}\omega}{2\pi} ~, \nonumber\\
&&\mathcal{P} \int_{-\infty}^{\infty} \Im[p _\mathbf{k}^\mathrm{TM}
(\omega)] \frac{\mathrm{d}\omega}{2\pi} =0 ~.
\end{eqnarray}
As already mentioned, $\Im[p_\mathbf{k}^\mathrm{TM}]$
has a $1/\omega$ singularity at the origin from the hyperbolic cotangent so
that the proper definition of the last equation above has to be
understood as a Cauchy's principal value $\mathcal{P}$. For the
other integrals, the functions are regular at the origin and the
principal value is not needed. For a singularity at a point $c$
in the domain of integration $\left[a,b\right]$, Cauchy's principal
value, represented by the symbol $\mathcal{P}$, is defined
as
\begin{equation}
  \label{eq:CauchyPV}
  \mathcal{P}\int_{a}^{b}\varphi (\omega) \text{d}\omega=
  \lim_{\epsilon \to 0^{+}}\left[\int_{a}^{c-\epsilon}+\int_{c+\epsilon}^{b}\right]
  \varphi(\omega)\text{d}\omega
\end{equation}

Applying Cauchy's residue theorem to the function $p
_\mathbf{k}^\varsigma (z)$ over the contours depicted in
Figs.~\ref{fig:contourDrude}, we rewrite \eqref{eq:pressure} in
terms of residues at the Matsubara poles $\imag\xi_n$
\begin{eqnarray}
\label{eq:pressureDrude} P = \sum_\mathbf{k} && \left(
\sum_{n=1}^\infty \underset{\imag\xi_n}{\mathrm{Res}} \left(\imag
p _\mathbf{k}^\mathrm{TE}(z) \right) \right. \\
&& \left. + \frac12 \underset{\imag\xi_0}{\mathrm{Res}} \left(\imag
p _\mathbf{k}^\mathrm{TM}(z) \right) + \sum_{n=1}^\infty
\underset{\imag\xi_n}{\mathrm{Res}} \left(\imag p
_\mathbf{k}^\mathrm{TM}(z) \right)  \right) ~. \nonumber
\end{eqnarray}
Substituting for the values of the residues in the last expression
leads to the final expression
\begin{eqnarray}
\label{eq:cauchyDrude} P=-2k_\mathrm{B}T \sum_\mathbf{k}
\sum_{\varsigma,n}^{\prime\prime} \kappa_n ~
f_\mathbf{k}^\varsigma[\imag \xi_n] ~,
\end{eqnarray}
with the double primed sum defined to match \eqref{eq:pressureDrude}
:
\begin{eqnarray}
\sum_{\varsigma,n}^{\prime\prime} \varphi^\varsigma(n) \equiv
\sum_{n=1}^\infty \varphi^\mathrm{TE}(n) + \sum_n^\prime
\varphi^\mathrm{TM}(n) ~. \label{eq:doubleprimedsum}
\end{eqnarray}
As there is no pole at $\imag\xi_0$ for the TE contribution, this
matches perfectly the expression \eqref{eq:pressureLM} for the
Casimir force between two thick metallic slabs described by the
Drude model. In the next section, we will go through the same
derivation for the plasma model and find an expression looking like
\eqref{eq:cauchyDrude} but differing from \eqref{eq:pressureLM}.

At this point, it is worth emphasizing a few points which have
played a role in the derivation of \eqref{eq:cauchyDrude}. First, we
have assumed the function $p _\mathbf{k}^\varsigma (z)$ to be
meromorphic in the domain enclosed by the contour $\mathcal{C}$.
This means in particular that a finite number of isolated
singularities lie in this domain, so that the temperature $T$ must
be strictly positive. Then, the contour has to be closed at infinity
with a vanishingly small contribution of the closing half-circle of
radius $1/\eta$. This is possible thanks to the so-called
transparency condition at high frequencies~\cite{Jaekel1991}, which
eliminates the possibility of considering perfect reflectors.
Finally, the Matsubara pole at the origin $\imag\xi_0=0$ must remain
isolated in order to be able to define a residue. This entails that
the Drude dissipation parameter $\gamma$ has to be strictly
positive, so that the poles associated with the Foucault currents
remain at finite distance from $\imag\xi_0$ (more discussions on
this point in the next section). Note also that branch cuts due to
$k_{z}$ starting at $\pm c\vert\mathbf{k}\vert$ approach $z_0$ if
$\vert\mathbf{k}\vert$ is not strictly positive. However, the
contribution from the single point $\vert\mathbf{k}\vert=0$ has a
null weight in the integral over $\mathbf{k}$ and does not contribute to
the final expression of the Casimir pressure.


\section{Motion of poles to the real axis} 
\label{sec:motionofpoles}

In the next section, we will discuss the plasma model which
corresponds to setting $\gamma=0$ in the Drude model permittivity
\eqref{eq:drudemodel}. It is commonly thought that this limiting
case, with a real permittivity function, is easier to handle than
the general case. We will show that this is not so, for reasons
which can be understood qualitatively from the remarks at the end of
the previous section. When moving from the Drude to the plasma
model, the poles of the functions $f_\mathbf{k}^\varsigma$ which
were lying strictly in the lower-half of the complex plane approach
the real axis when $\gamma\to0$ and touch it when $\gamma=0$. It
follows that the application of Cauchy's residue theorem is much
more delicate for $\gamma=0$ than for $\gamma>0$.

We first discuss the motion of the poles of the functions
$f_\mathbf{k}^\varsigma$, which correspond to propagating
Fabry-Perot or surface plasmon modes. For the Drude model, we denote
by $z_m^\varsigma=\omega_m^\varsigma-\imag \epsilon_m^\varsigma$ the
positions of these poles, which also depend on
$\vert\mathbf{k}\vert$. When $\gamma \to 0$, the poles approach the
real axis with $\epsilon_m^\varsigma \to 0$.
There is a finite number of such poles and they lie in the interval
$0>\omega_m^\varsigma>\sqrt{\omega_\mathrm{p}^{2}+c^{2}\mathbf{k}^{2}}$
. For each of these
isolated single poles, we may introduce a punctured disk in which
the function has a Laurent series expansion
\begin{eqnarray}
\label{eq:laurent} &&p _\mathbf{k}^\varsigma (z) =
\sum_{\ell=-1}^\infty a_\ell (z-z_m^\varsigma)^\ell \equiv
\frac{\hat{p}_m^\varsigma(z)}{z-z_m^\varsigma} ~, \nonumber \\
&&z\in\mathcal{D}_m^\varsigma \quad:\quad 0< \vert z-z_m^\varsigma
\vert < R_m^\varsigma ~.
\end{eqnarray}
The radius $R_m^\varsigma$ of the disk is chosen so that $p
_\mathbf{k}^\varsigma$ is holomorphic in $\mathcal{D}$. The first
term in the Laurent series \eqref{eq:laurent} is related to the
residue at the pole
\begin{equation}
\label{eq:residue} \underset{z_m^\varsigma}{\mathrm{Res}} \left(p
_\mathbf{k}^\varsigma\right) = a_{-1} =
\hat{p}_m^\varsigma(z_m^\varsigma) ~.
\end{equation}
We will use these properties in the next section in order to
calculate the Casimir pressure.

We then consider the poles associated with the Foucault currents
which have already been studied for the Drude
model~\cite{Intravaia2009,Intravaia2010}, which poles lie in the
interval $\mathcal{I}$ on the lower half of the imaginary axis
\begin{equation}
\label{eq:foucaultinterval} \mathcal{I} \quad:\quad
-\tilde{\gamma}<\xi<-\gamma ~.
\end{equation}
Here $\tilde{\gamma}$ is a positive real number defined as the real
root of the cubic equation $\left(x^2+c^2\mathbf{k}^2 +
\omega_\mathrm{p}^2\right) x = \left(x^2 + c^2\mathbf{k}^2\right)
\gamma$. In the limit $\gamma\to0$, it is simply $\tilde\gamma
\simeq \gamma c^2\mathbf{k}^2 / \left( c^2\mathbf{k}^2 +
\omega_\mathrm{p}^2 \right)$. This entails that the Foucault poles
remain at finite distance from the origin, except for
$\vert\mathbf{k}\vert=0$, which, as already said, has a null weight
in the sum over $\mathbf{k}$.

\begin{figure}[htbp]
  \begin{center}
    \includegraphics[width=2.0in]{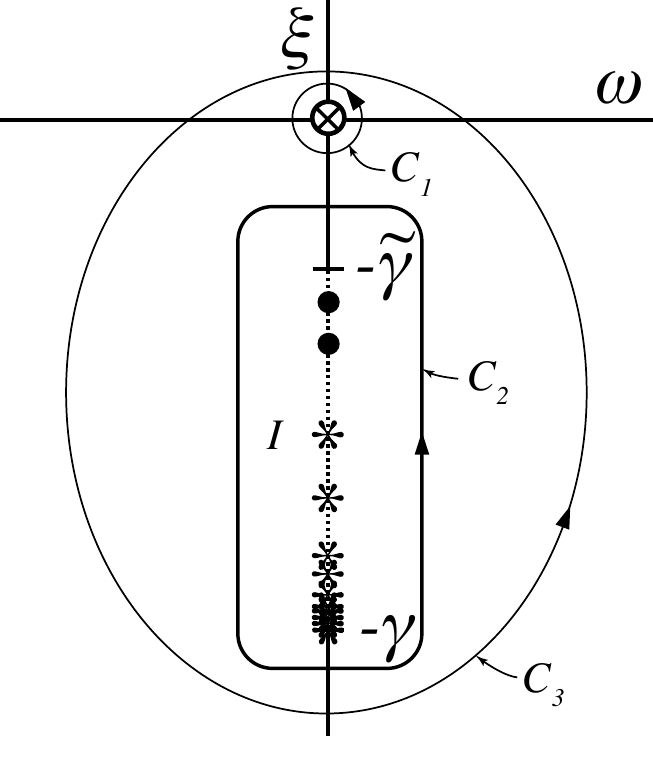}
  \end{center}
\caption{Schematic representation of the poles and zeros of the
function $p _\mathbf{k}^\mathrm{TE}$ in the vicinity of the origin
for the Drude model. A zero at the origin is indicated by a cross,
while other poles and zeros in the interval $\mathcal{I}$ are shown
as black dots and asterisks (interpretation of these symbols in
Fig.~\ref{fig:appearancePoles}). } \label{fig:zp}
\end{figure}

Fig.~\ref{fig:zp} shows a schematic representation of the poles and
zeros of the function $p _\mathbf{k}^\mathrm{TE}$ in the vicinity of
the origin for the Drude model. There is one pole due to $C$ and a
double zero due to $f_\mathbf{k}^\mathrm{TE}$, which combine in a
zero indicated by the cross at the origin. There then exist a
countable infinity of poles and zeros lying in the interval
$\mathcal{I}$. Finally, we have to take care of the Matsubara poles
which might belong to this interval depending on the respective
values of $\gamma$ and $T$. For the case of gold at room
temperature, $\xi_1$ ($\sim160$meV) is larger than $\gamma$
($\sim35$meV), so that we will consider in the following that there
is no Matsubara pole in the interval $\mathcal{I}$. Should this not
be the case ($\gamma$ larger than for gold), the Matsubara poles
could easily be counted as their positions depend only on $T$, and
not on the parameters determining the Foucault poles and zeros.

If the poles in the interval $\mathcal{I}$ were isolated from each
other, the functions $p _\mathbf{k}^\mathrm{TE}$ or $f
_\mathbf{k}^\mathrm{TE}$ would be meromorphic and it would be
possible to use Cauchy's argument principle to obtain information on
the number of their poles and zeros. As a matter of fact, the number
\begin{equation}
\label{eq:argument} N \equiv \oint_\mathcal{C}
\frac{f^\prime(z)}{f(z)} \frac{\mathrm{d}z}{2 \imag\pi}
\end{equation}
would be $N=Z-P$ with $Z$ and $P$ the numbers of zeros and poles of
$f$ enclosed in $\mathcal{C}$. In the following, we show how to use
an extension of the argument principle to obtain the same kind of
information even though the functions $p _\mathbf{k}^\mathrm{TE}$ or
$f _\mathbf{k}^\mathrm{TE}$ are not meromorphic while the associated
$Z$ and $P$ are both infinite.

Numerical evaluations of the number $N$ defined by
\eqref{eq:argument}, for the functions $p _\mathbf{k}^\mathrm{TE}$
and $f _\mathbf{k}^\mathrm{TE}$ and the contour $\mathcal{C}_1$
shown on Fig.~\ref{fig:zp}, confirms the already known fact that
$N=1$ for the function $p _\mathbf{k}^\mathrm{TE}$ on the contour
surrounding the origin, with this number being the sum of two zeros
for the function $f _\mathbf{k}^\mathrm{TE}$ and one pole for the
function $C$. A new result $N=-2$ is obtained for the contour
$\mathcal{C}_2$ enclosing the interval $\mathcal{I}$, which means
that there are two more poles than zeros there. Finally, for the
contour $\mathcal{C}_3$ enclosing both the origin and $\mathcal{I}$,
we obtain $N=-1$, that is, the sum of results for $\mathcal{C}_1$
and $\mathcal{C}_2$.

\begin{figure}[htbp]
  \begin{center}
    \includegraphics[width=3.2in]{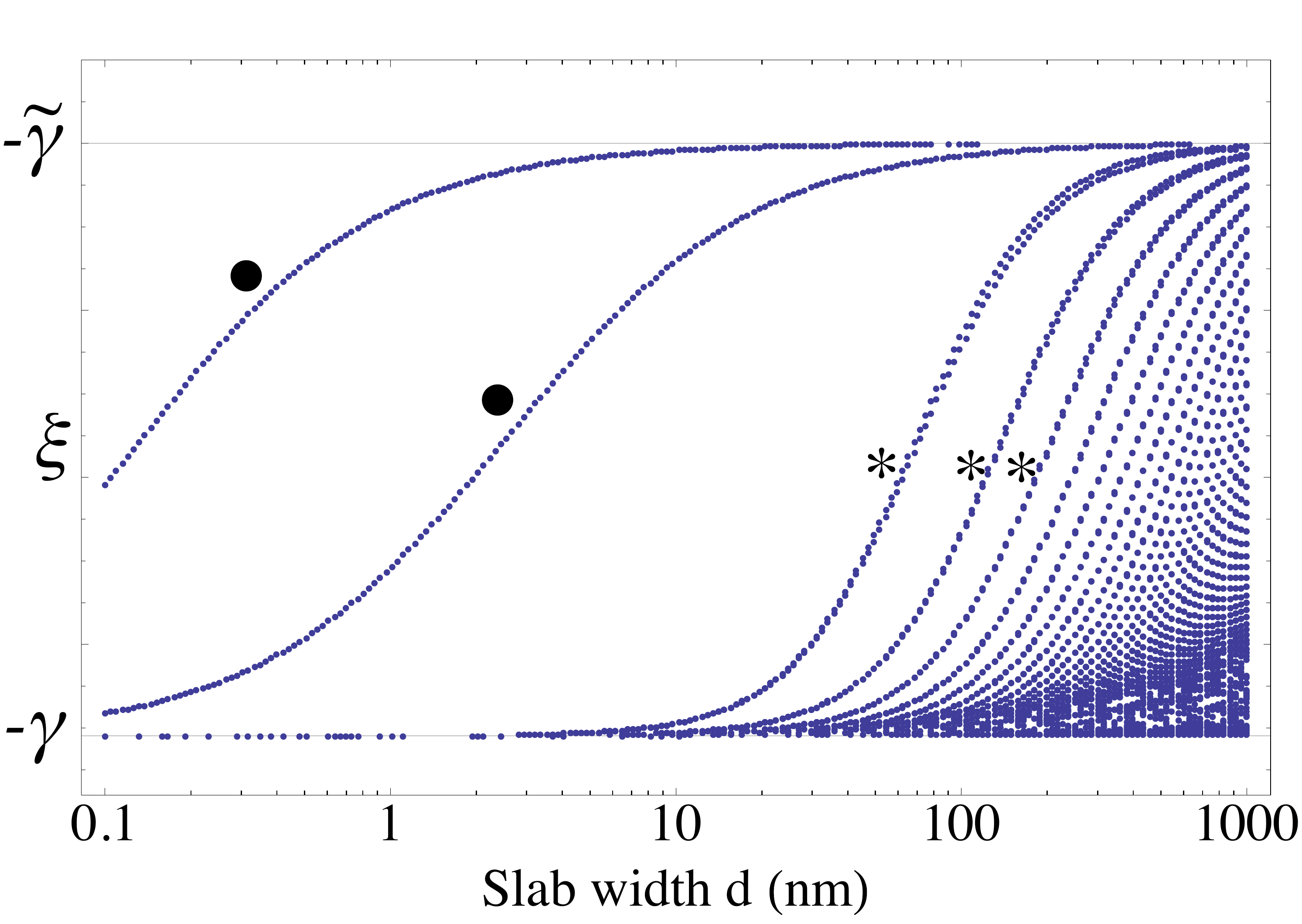}
  \end{center}
\caption{Positions on the imaginary axis of the poles and zeros of
the function $p_\mathbf{k}^\mathrm{TE}$ in the interval
$\mathcal{I}$ as a function of the width $d$ of the slabs. Each
trajectory marked with a dot follows one single pole whereas each
trajectory with an asterisk follows a group of one double zero and
two simple poles.}
  \label{fig:appearancePoles}
\end{figure}

The preceding discussion shows that, even though Cauchy's argument
principle cannot be applied in its common form because $Z$ and
$P$ are both infinite, the value of $N=Z-P$ is still well-defined.
The same assertion can be understood through a different reasoning,
which follows the variation of the poles and zeros when the width
$d$ of the slabs is changed. For very thin slabs $d \to 0$, there
are essentially two poles whose trajectories as a function of $d$
are marked with dots in Fig.~\ref{fig:appearancePoles}. When the
value of $d$ is increased, we see the appearance of the infinite
number of poles and zeros. Those poles and zeros emerge from the
branch point $\xi=-\gamma$ as groups of one zero of order two and
two poles of order one, each group marked with an asterisk in
Fig.~\ref{fig:appearancePoles}. These poles and zeros then fill the
whole interval $\mathcal{I}$ when $d\to\infty$. This reasoning
explains why we can consider that there are two more Foucault poles
than zeros there.

In the limit $\gamma\to0$, all the poles and zeros converge to the
origin, where they collapse into what can be considered as a single
pole, according to the discussion just presented in terms of
Cauchy's argument principle. Again these properties will play a key
role in the next section for the calculation of the Casimir
pressure.

For the sake of completeness, we also discuss the TM
case. The function $f_{\mathbf{k}}^{\text{TM}}$ behaves as $z^{0}$
near the origin, and the function $p_{\mathbf{k}}^{\text{TM}}$ has a
pole at the origin due to $C$. In the interval $\mathcal{I}$,
$p_{\mathbf{k}}^{\text{TM}}$ possesses an equal number of poles and
zeros so that we obtain $N=0$ on the contour $\mathcal{C}_{2}$. The
result $N=-1$ on the contour $\mathcal{C}_{3}$ is common to both
polarizations. The difference in the number of poles and zeros in
the interval $\mathcal{I}$ for TE and TM leads to a fundamental
difference at the origin in the spectral density
$\Re[p_{\mathbf{k}}^{\varsigma}(\omega)]$. These points will be
discussed in more detail in the next section.


\section{The plasma model} 
\label{sec:plasmamodel}

We come now to the calculation of the Casimir pressure for mirrors
described by the plasma model and, in particular, we carefully
discuss the derivation of the Lifshitz-Matsubara sum formula
starting from the integral \eqref{eq:pressure} over real
frequencies. The application of Cauchy's residue theorem is much
more delicate for $\gamma=0$ than for $\gamma>0$, because the
integrand $p _\mathbf{k}^\varsigma$ has now to be understood in
terms of distributions.

We first consider the simplest case of an isolated single pole
$z_m^\varsigma = \omega_m^\varsigma - \imag \epsilon_m^\varsigma$ of
the function $f_\mathbf{k}^\varsigma$ corresponding to a propagating
Fabry-Perot or plasmonic mode. This single mode lies below the real
axis for $\gamma>0$ and comes to touch the real axis with
$\epsilon_m\to0$ when $\gamma\to0$. It is thus clear that the
integrand $p _\mathbf{k}^\varsigma (\omega)$ in \eqref{eq:pressure}
contains a Dirac delta distribution associated with this pole. The
associated contribution is easily evaluated by applying the
so-called Sokhotsky's formula~\cite{Vladimirov1971}
\begin{eqnarray}
\label{eq:sokhotsky} \lim_{\epsilon \rightarrow 0^+} \int_a^b
\frac{\varphi(\omega)}{\omega-\omega_m \mp \imag \epsilon} \text{d}\omega&=&
\pm i \pi \varphi(\omega_m) \\
&+& \mathcal{P} \int_a^b \frac{\varphi(\omega)}{\omega-\omega_m} \text{d}\omega
~.\nonumber
\end{eqnarray}
As the pole is isolated from other ones, we can consider values of
$\gamma$ small enough so that the disk $\mathcal{D}_m^\varsigma$
introduced in \eqref{eq:laurent} includes a segment
$\mathcal{S}_m^\varsigma$ covering the vicinity of the pole on the
real axis. We then apply Sokhotsky's formula \eqref{eq:sokhotsky} to
the function \eqref{eq:laurent} on this segment to obtain
\begin{equation}
\int_{\mathcal{S}_m^\varsigma}
\frac{\hat{p}_m^\varsigma(z)}{z-z_m^\varsigma}
\frac{\mathrm{d}\omega}{2\pi} = \mathcal{P}
\int_{\mathcal{S}_m^\varsigma}
\frac{\hat{p}_m^\varsigma(z)}{z-z_m^\varsigma}
\frac{\mathrm{d}\omega}{2\pi} - \frac\imag2
\hat{p}_m^\varsigma(z_m^\varsigma) ~.
\end{equation}
The real part of the last equation is also
\begin{equation}
\label{eq:sokhotsky2} \int_{\mathcal{S}_m^\varsigma} \Re[p
_{\mathbf{k}}^\varsigma (\omega)] \frac{\mathrm{d}\omega}{2\pi} =
\mathcal{P} \int_{\mathcal{S}_m^\varsigma} \Re[p
_{\mathbf{k}}^\varsigma (\omega)] \frac{\mathrm{d}\omega}{2\pi} +
\frac12 \Im[\hat{p}_m^\varsigma(z_m^\varsigma)] ~,
\end{equation}
and the last term in \eqref{eq:sokhotsky2} is related to the residue
\eqref{eq:residue}. This residue is easily seen to be purely
imaginary for ordinary as well as evanescent modes.

\begin{figure}[htbp]
  \begin{center}
    \includegraphics[width=3.2in]{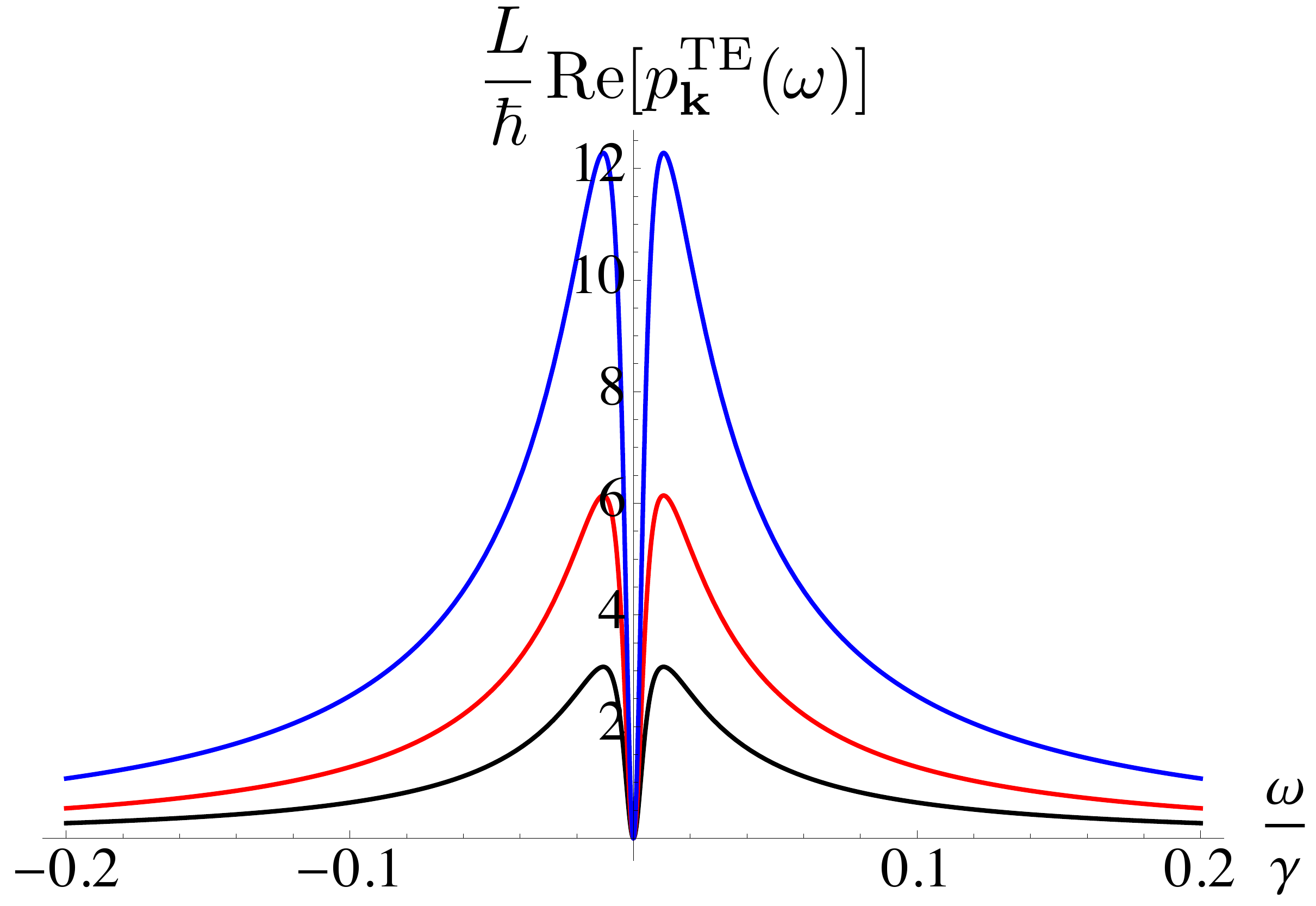}
    \includegraphics[width=3.2in]{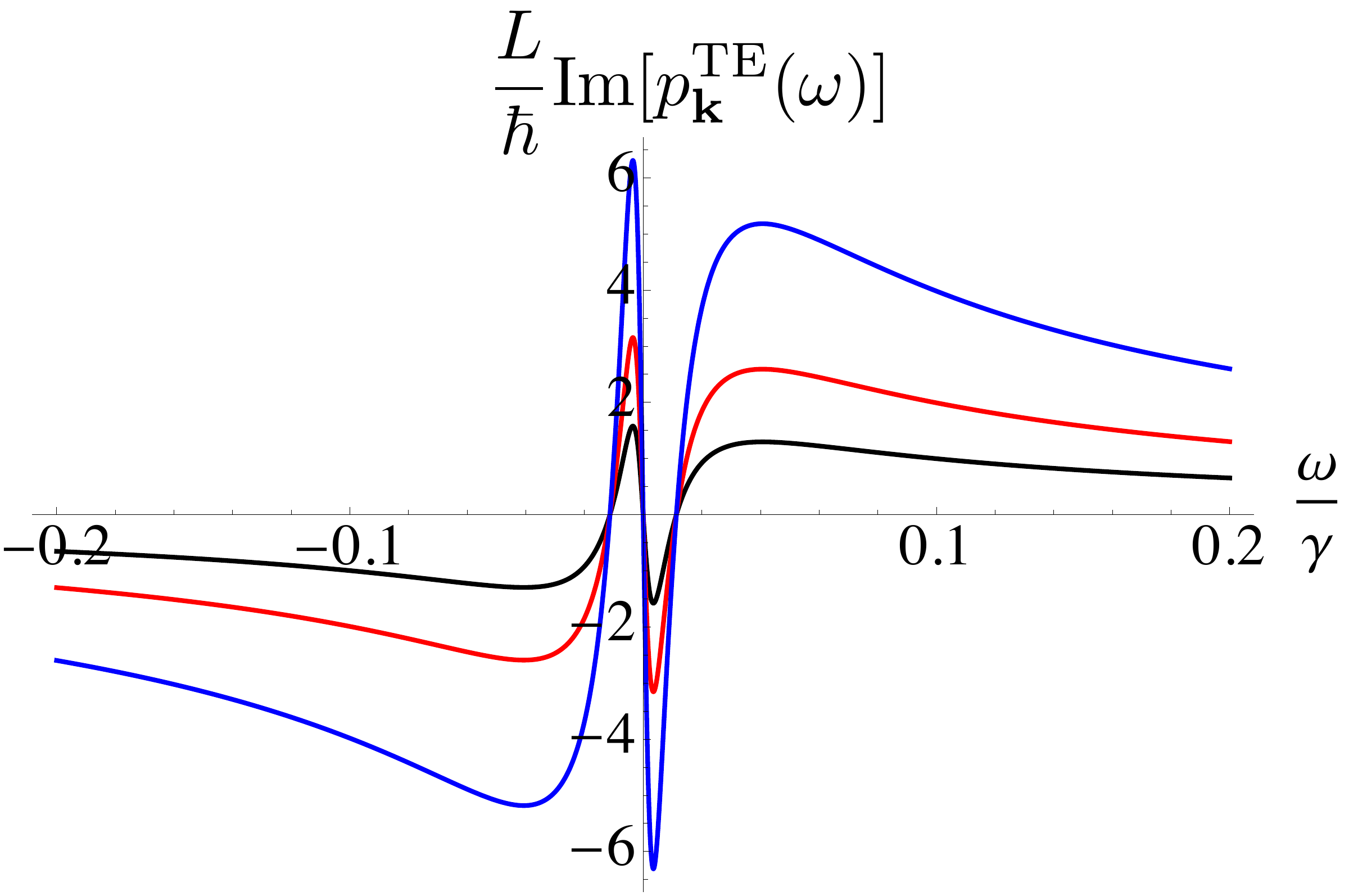}
  \end{center}
\caption{Real (top plot) and imaginary part (bottom plot) of the dimensionless
$\frac{L}{\hbar}p_\mathbf{k}^\mathrm{TE}$ drawn in the vicinity of the origin on the
real axis, as a function of the dimensionless parameter
$\omega/\gamma$. The curves on each plot with increasing maxima
correspond to values of $\gamma$ being 1/4 (blue), 1/2 (red) and 1
(black), respectively, of that of gold. [Colors online]}
\label{fig:TE-drudePlasma}
\end{figure}

We then come to the case of modes associated to Foucault current
which deserve a specific treatment, as already discussed in
section~\ref{sec:motionofpoles}. We use the property shown there
that all the poles and zeros lying in the vicinity of the origin for
the Drude model (see Fig~\ref{fig:zp}) converge in the limit
$\gamma\to0$ to a single pole touching the origin. This idea is
confirmed by the drawing in Fig.~\ref{fig:TE-drudePlasma} of the
real (top plot) and imaginary (bottom plot) parts of the function $p
_\mathbf{k}^\mathrm{TE}$ as a function of $\omega$ in the vicinity
of the origin. For each plot, the bottom curve corresponds to a
calculation with the parameters chosen to match gold (with
$L=250$ nm), while the middle and top curves correspond to values of
$\gamma$ divided respectively by factors 2 and 4.

\begin{figure}[htbp]
  \begin{center}
    \includegraphics[width=3.2in]{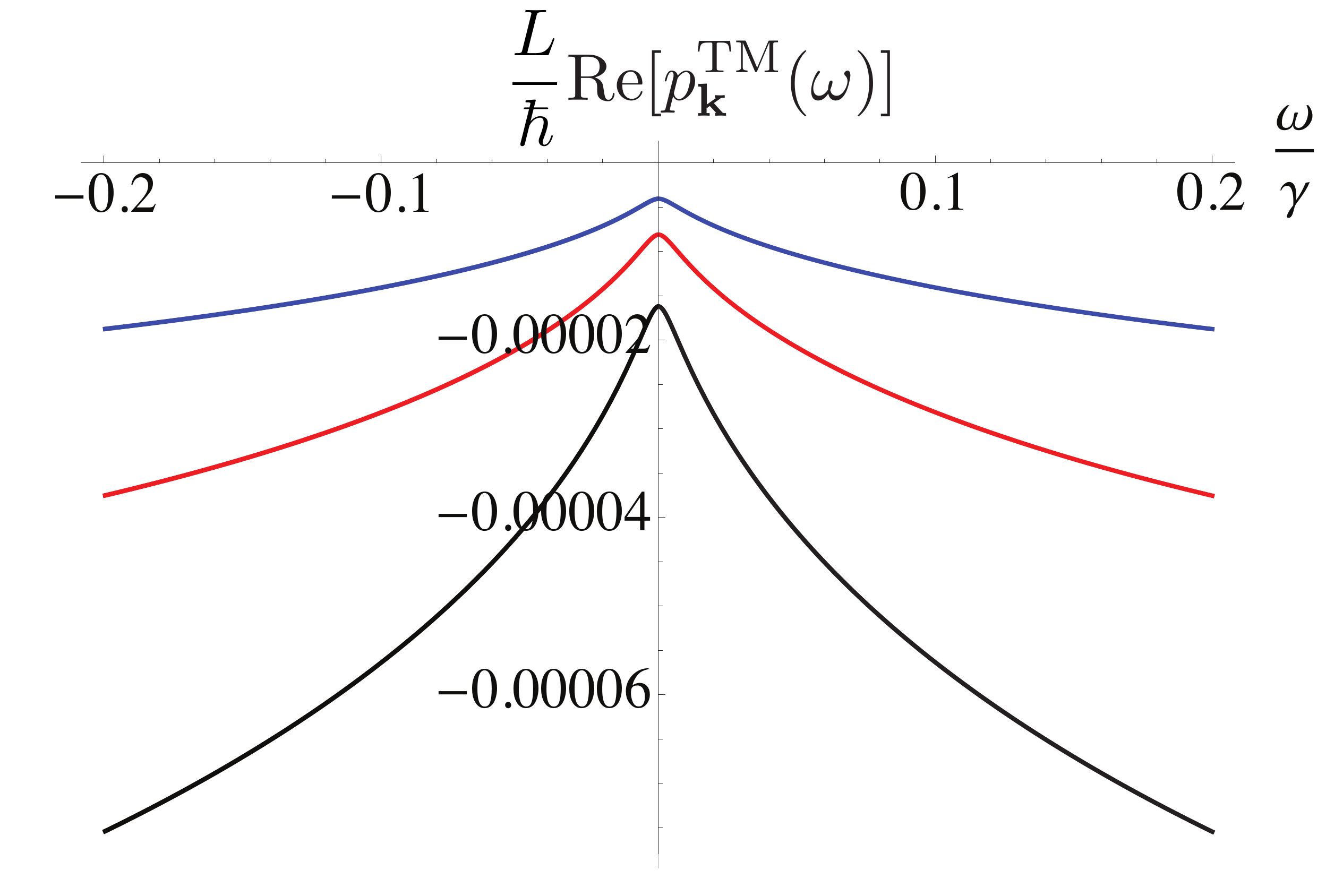}
  \end{center}
\caption{Real part of the dimensionless $\frac{L}{\hbar}p _\mathbf{k}^\mathrm{TM}$
drawn in the vicinity of the origin on the real axis. Same conventions as for
Fig.~\ref{fig:TE-drudePlasma}. [Colors online]}
\label{fig:TM-drudePlasma}
\end{figure}

The plots on Fig.~\ref{fig:TE-drudePlasma} clearly
show that the middle and top curves are identical to the bottom
curve multiplied by two and four. As the curves are drawn as a
function of the dimensionless quantity $\omega/\gamma$, their
integrals tend to a finite limit when $\gamma\to0$. This scaling
property implies that the function $p _\mathbf{k}^\mathrm{TE}$ has a
singularity at the origin, with its real part containing the
equivalent of a Dirac delta function $\delta(\omega)$ when
$\gamma\to0$. This Dirac delta function can be treated by
Sokhotsky's formula as in the already discussed case of isolated
poles. The situation is clearly different for the TM polarization,
as shown by the plots on Fig.~\ref{fig:TM-drudePlasma} (same
conventions as for Fig.~\ref{fig:TE-drudePlasma}). The function
$\Re[p _\mathbf{k}^\mathrm{TM}]$ tends to 0 when $\gamma\to0$,
so that there is no singularity left at the origin. The
difference between TE and TM cases is directly related to the
counting of poles and modes in the preceding section.

After this discussion, Sokhotsky's formula now allows us to write
the following relations which are the counterpart of
\eqref{eq:parityDrude} for the plasma model for the contribution to
the Casimir pressure for given values of $\mathbf{k}$ and $\varsigma$
\begin{eqnarray}
\label{eq:parityPlasma} \int_{-\infty}^{\infty} \Re[p
_\mathbf{k}^\varsigma (\omega)] \frac{\mathrm{d}\omega}{2\pi} &=&
\mathcal{P} \int_{-\infty}^{\infty} \Re[p _\mathbf{k}^\varsigma
(\omega)] \frac{\mathrm{d}\omega}{2\pi} \nonumber \\
&+&\frac12 \sum_m \underset{\omega_m^\varsigma}{\mathrm{Res}}
\left(-\imag p _\mathbf{k}^\varsigma(z)\right) ~, \\
\mathcal{P}\int_{-\infty}^{\infty} \Im[p _\mathbf{k}^\varsigma
(\omega)] \frac{\mathrm{d}\omega}{2\pi} &=&0 ~. \nonumber
\end{eqnarray}
$\omega_m^\varsigma$ are the poles on the real axis of $p
_\mathbf{k}^\varsigma$, which come as pairs symmetrically located
with respect to the imaginary axis, except for those sitting at the
origin. For the TE polarization, the poles are labeled by integers
$m\in\mathds{Z}$. The number $m=0$ corresponds to the pole at
$\omega_0=0$ which collects all the poles and zeros in the vicinity
of the origin, as discussed in section~\ref{sec:motionofpoles},
while the non-zero integers with opposite signs correspond to poles
symmetrically located with respect to the imaginary axis. For the TM
polarization, there is no pole of $f_\mathbf{k}^\mathrm{TM}$ on the
imaginary axis, so that the poles are labeled by integers
$m\in\mathds{Z}^\ast$ (\textit{i.e.} $m\in\mathds{Z}$ and $m\neq0$).
To repeat, in eq.~(\ref{eq:parityPlasma}) the
Cauchy's principal value $\mathcal{P}$ is taken at
each singularity of the function $\Im[p_{\mathbf{k}}^{\varsigma}(\omega)]$ while the
last term originating from the application of Sokhotsky's formula counts
the contributions of Dirac delta distributions in $\Re[p_{\mathbf{k}}^{\varsigma}(\omega)]$.

\begin{figure}[htbp]
  \begin{center}
    \includegraphics[width=3.2in]{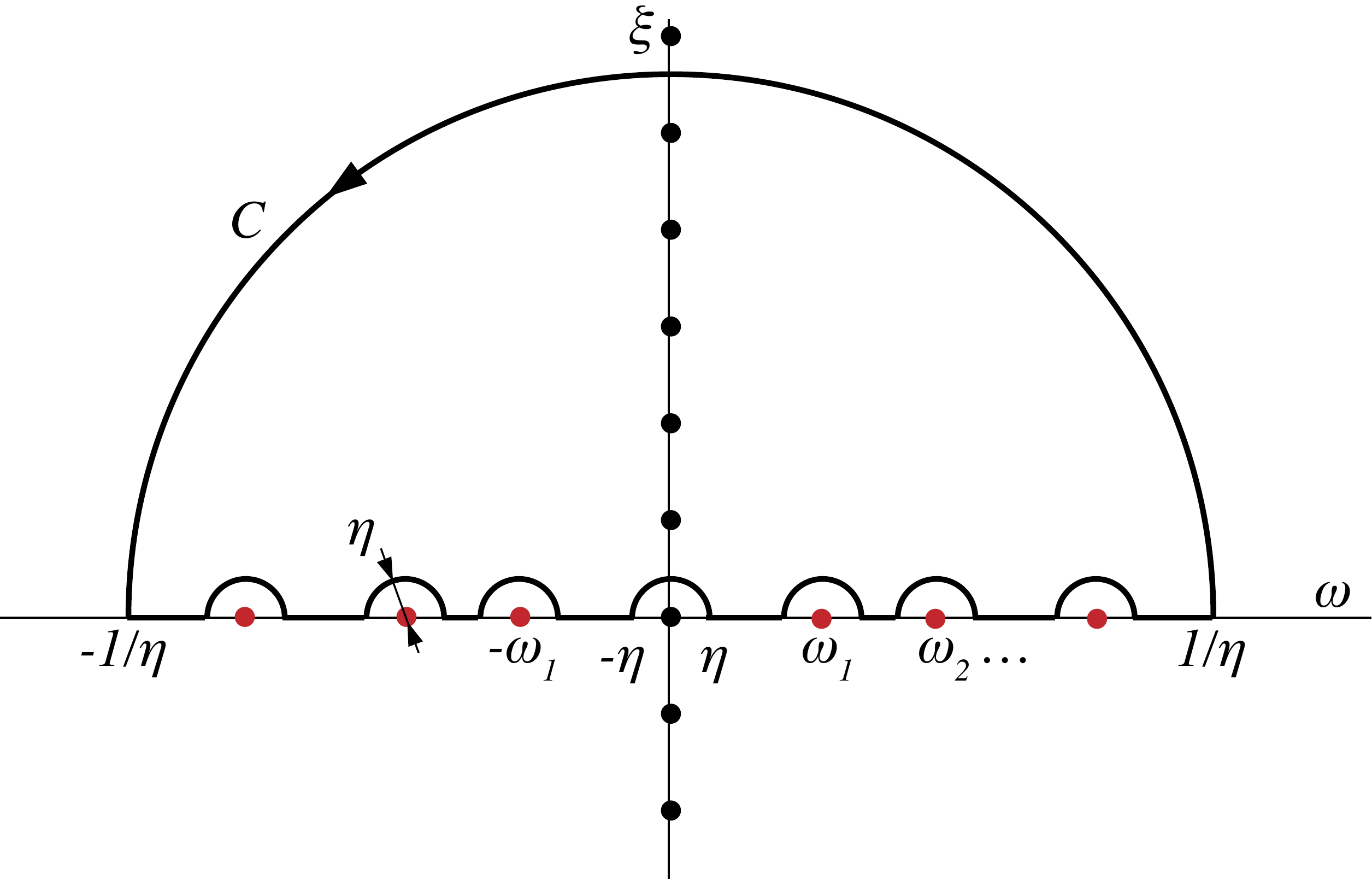}
  \end{center}
\caption{Contour used for the application of Cauchy's residue
theorem for metallic slabs described by the plasma model. The finite
number of poles of the function $f_\mathbf{k}^\varsigma$ are represented
as red dots and those due to $C(z)$ by black dots. [Colors online]}
  \label{fig:contourPlasma}
\end{figure}

As in section \ref{sec:drudemodel}, we now proceed to the derivation
of the Lifshitz-Matsubara sum formula by applying Cauchy's residue
theorem to the function $p _\mathbf{k}^\varsigma (z)$ over the
contour shown on Fig.~\ref{fig:contourPlasma}, now used for both
polarizations. A subtlety arises here, as the branch cut due to
$k_z$ lying on the real axis for $\omega>c\vert\mathbf{k}\vert$ may
prevent us to define the punctured disk $\mathcal{D}
_\mathbf{k}^\varsigma$ as previously. A way out of this difficulty
is to define a cut with indentations around the poles as was done
in~\cite{Nesterenko2012}. Cauchy's residue theorem on the contour
$\mathcal{C}$ then leads, for both polarizations, to
\begin{eqnarray}
\label{eq:cauchyPlasma} \mathcal{P} \int_{-\infty}^\infty p
_\mathbf{k}^\varsigma (\omega) \frac{\mathrm{d}\omega}{2\pi} &=&
\frac12 \sum_{m\neq0} \underset{\omega_m^\varsigma}{\mathrm{Res}}
\left( \imag
p _\mathbf{k}^\varsigma (z) \right) \\
&+& \frac12 \underset{\imag \xi_0}{\mathrm{Res}} \left( \imag p
_\mathbf{k}^\varsigma (z) \right) + \sum_{n=1}^\infty
\underset{\imag \xi_n^\varsigma}{\mathrm{Res}} \left( \imag p
_\mathbf{k}^\varsigma (z) \right) ~, \nonumber
\end{eqnarray}
where Cauchy's principal value is again taken at all modes on the
real axis, including $\omega_0^\mathrm{TE}=0$.

Collecting these results with those in \eqref{eq:parityPlasma}, we
obtain the Casimir pressure between thick metallic slabs described
by the plasma model
\begin{eqnarray}
\label{eq:finalPlasma} P&=&-2k_\mathrm{B}T \sum_\mathbf{k} \left(
\sum_\varsigma \sum_n^\prime \kappa_n f_\mathbf{k}^\varsigma (\imag
\xi_n)
- \frac12 \kappa_0 f_\mathbf{k}^\mathrm{TE}(0) \right)\nonumber \\
&=&-2k_\mathrm{B}T \sum_\mathbf{k} \sum_{\varsigma,n}^{\prime\prime}
\kappa_n f_\mathbf{k}^\varsigma (\imag \xi_n)  ~,
\end{eqnarray}
where the primed and double primed sum symbols are respectively
defined in \eqref{eq:primedsum} and \eqref{eq:doubleprimedsum}. It
turns out that \eqref{eq:finalPlasma} has the same form as the final
expression \eqref{eq:cauchyDrude} obtained for the Drude model in
section~\ref{sec:drudemodel}. In contrast to the Drude case however,
the expression \eqref{eq:finalPlasma} is no longer identical to the
commonly used \eqref{eq:pressureLM}. This is obvious in the first
line of \eqref{eq:finalPlasma} where the first term matches
\eqref{eq:pressureLM} whereas the last term, comes to cancel the TE
contribution at the Matsubara frequency $\xi_0$. This cancellation
is the main result of the present paper, where it has been deduced
through a careful application of Cauchy's residue theorem to the
function $p _\mathbf{k}^\varsigma$ appearing in the integral
expression of the Casimir pressure.


\section{Discussion} 
\label{sec:discussion}

The Casimir pressure between thick slabs described by a local
dielectric function was derived by Lifshitz~\cite{Lifshitz1956} by
using the fluctuations-dissipation theorem~\cite{Callen1951} and
then confirmed by Dzyaloshinskii, Lifshitz and
Pitaevskii~\cite{Dzyaloshinskii1961}. The original derivation by
Lifshitz, developed in the spirit of Rytov's method with
fluctuations originating from matter~\cite{Rytov1953}, is perfectly
correct for the case of thick slabs made of dissipative
media~\cite{Polder1971,Lifshitz1980,Rytov1989,Antezza2008}. In such
a method, the plasma model can only be considered as the limit
$\gamma\to0$ of the dissipative Drude model.

In the present paper, we have used the derivation of the Casimir
pressure as the result of vacuum and thermal radiation pressure on
the two mirrors~\cite{Jaekel1991}. This approach is valid for lossy
as well as lossless mirrors~\cite{Genet2003,Lambrecht2006} while
reproducing the Lifshitz formula for reflection amplitudes deduced
from Fresnel equations (eq.\eqref{eq:pressure} in the present
paper). The remark in the preceding paragraph thus has crucial
consequences for the derivation of the Lifshitz-Matsubara sum
formula (eq.~(5.2) in~\cite{Lifshitz1956}, that is also
\eqref{eq:pressureLM} with the notations of the present paper).

We have confirmed the validity of this formula for dissipative
metals as well as dielectrics, for which
$r_\mathbf{k}^\mathrm{TE}(0,k)$ vanishes (eq.\eqref{eq:pressureLM}
is equivalent to eq.\eqref{eq:cauchyDrude} in this case). This
formula shows the nice property of having a completely symmetrical
form for the two polarizations but it also leads to a discontinuity
in the calculated thermal Casimir pressure when going from a
dissipative model ($\gamma\neq0$) to a non-dissipative one
($\gamma=0$).

For thick metallic slabs described by the plasma model, we have
found the expression \eqref{eq:finalPlasma} for the Casimir
pressure, which is not identical to the one commonly used
\eqref{eq:pressureLM}. This follows from the fact that the Matsubara
pole for the TE mode at $\xi_0$ does not contribute for a lossless
plasma metal, in spite of a non-vanishing value of
$r_\mathbf{k}^\mathrm{TE}(0,k)$. As a consequence, the discontinuity
in the calculated thermal Casimir pressure between dissipative and
non-dissipative metals disappears. It has however to be acknowledged
that this result does not solve the discrepancy observed between
theory and some
experiments~\cite{Klimchitskaya2006,Brevik2006,Lambrecht2011}.

The interpretation of this result is that the contribution
to the pressure corresponding to the non-vanishing value of
$r_\mathbf{k}^\mathrm{TE}(0,k)$ is canceled by an additional
contribution originating from the collapse of all poles due
to the Foucault modes at the origin of the complex plane.
This has been proven by two different but
equivalent approaches, first by counting the poles and zeros
of the causal function $p _\mathbf{k}^\mathrm{TE}$ (\S~IV), and
then by examining the behavior in the vicinity of the origin
of the density $\Re [p _\mathbf{k}^\mathrm{TE}]$ (\S~V).
We have shown in the present paper that this contribution
of Foucault modes, usually ignored in calculations
of the Casimir pressure for a lossless plasma model, leaves
a finite contribution in the limit $\gamma\to0$, which is just
the difference between the common form of the Lifshitz-Matsubara
sum formula and its corrected expression.


\acknowledgments KAM thanks the Laboratoire Kastler Brossel for
their hospitality during the period of this work, the Simons
Foundation and the Julian Schwinger Foundation for financial
support. The authors thank the CASIMIR network
(www.casimir-network.org). We thank Prachi Parashar for helpful
discussions.


\newcommand{\REVIEW}[4]{\textit{#1} \textbf{#2} #3 (#4)}
\newcommand{\Review}[1]{\textit{#1}}
\newcommand{\Volume}[1]{\textbf{#1}}
\newcommand{\Book}[1]{\textit{#1}}
\newcommand{\Eprint}[1]{\textsf{#1}}
\def\etal{\textit{et al}}


\begin{thebibliography}{999.}

\bibitem{Casimir1948}
H.B.G. Casimir,
\Review{Proc. K. Ned. Akad. Wet. (Phys.)} \Volume{51} 79 (1948).

\bibitem{Milton2001}
K.A. Milton,
\Book{The Casimir effect, physical manifestation of zero-point energy}
(World Scientific, 2001).

\bibitem{Klimchitskaya2009}
G.L. Klimchitskaya, U. Mohideen and V.M. Mostepanenko,
\Review{Rev. Mod. Phys.} \Volume{81} 1827 (2009).

\bibitem{Lamoreaux2011}
S. Lamoreaux, in \Book{Casimir physics},
eds. D.A.R. Dalvit \etal,
Lecture Notes in Physics \Volume{834} (Springer-Verlag, 2011) p.219.

\bibitem{Capasso2011}
F. Capasso, J.N. Munday, and H.B. Chan, in \Book{Casimir physics},
eds. D.A.R. Dalvit \etal,
Lecture Notes in Physics \Volume{834} (Springer-Verlag, 2011) p.249.

\bibitem{Decca2011}
R. Decca, V. Aksyuk, and D. L\'opez, in \Book{Casimir physics},
eds. D.A.R. Dalvit \etal,
Lecture Notes in Physics \Volume{834} (Springer-Verlag, 2011) p.287.

\bibitem{Klimchitskaya2006}
G.L. Klimchitskaya and V.M. Mostepanenko,
\Review{Contemp. Phys.} \Volume{47} 131 (2006).

\bibitem{Brevik2006}
I. Brevik, S.\AA. Ellingsen and K. Milton,
\Review{New J. Phys.} \Volume{8} 236 (2006).

\bibitem{Lambrecht2011}
A. Lambrecht, A. Canaguier-Durand, R. Gu\'erout and S. Reynaud,
in \Book{Casimir physics},
eds. D.A.R. Dalvit \etal,
Lecture Notes in Physics \Volume{834} (Springer-Verlag, 2011) p.97.

\bibitem{Lifshitz1956} E.M Lifshitz,
\Review{Sov. Phys. JETP} \Volume{2} 73 (1956).

\bibitem{Dzyaloshinskii1961}
I.E. Dzyaloshinskii, E.M. Lifshitz and L.P. Pitaevskii,
\Review{Sov. Phys. Uspekhi} \Volume{4} 153 (1961).

\bibitem{Jaekel1991}
M.-T. Jaekel and S. Reynaud,
\Review{J. Physique I} \Volume{1} 1395 (1991).

\bibitem{Lambrecht2000}
A. Lambrecht and S. Reynaud,
\Review{Euro. Phys. J.} \Volume{D 8} 309 (2000).

\bibitem{Svetovoy2008}
V.B. Svetovoy, P.J. van Zwol, G. Palasantzas and J.Th.M. De Hosson,
\Review{Phys. Rev. B} \Volume{77} 035439 (2008).

\bibitem{Bostrom2000}
M. Bostr\"om and B.E. Sernelius,
\Review{Phys. Rev. Lett.} \Volume{84} 4757 (2000).

\bibitem{Ingold2009}
G.-L. Ingold, A. Lambrecht and S. Reynaud,
\Review{Phys. Rev. E} \Volume{80} 041113 (2009).

\bibitem{Brevik2014}
I. Brevik and J.S. H{\o}ye,
\Review{Eur. J. Phys.} \Volume{35} 015012 (2014).

\bibitem{Decca2007prd}
R.S. Decca, D. L\'opez, E. Fischbach  \etal,
\Review{Phys. Rev. D} \Volume{75} 077101 (2007).

\bibitem{Decca2007epj}
R.S. Decca, D. L\'opez, E. Fischbach  \etal,
\Review{Eur. Phys. J.} \Volume{C 51} 963 (2007).

\bibitem{Chang2012}
C.-C. Chang, A. A. Banishev, R. Castillo-Garza \etal,
\Review{Phys. Rev. B} \Volume{85} 165443 (2012).

\bibitem{Sushkov2011}
A.O. Sushkov, W.J. Kim, D.A.R. Dalvit and S.K. Lamoreaux,
\Review{Nature Physics} \Volume{7} 230 (2011).

\bibitem{Milton2011}
K. Milton, \Review{Nature Physics} \Volume{7} 190 (2011).

\bibitem{Kim2010}
W.J. Kim, A.O. Sushkov, D.A.R. Dalvit and S.K. Lamoreaux,
\Review{Phys. Rev. A} \Volume{81} 022505 (2010).

\bibitem{Reynaud2014}
S. Reynaud and A. Lambrecht, in \Book{Quantum Optics and Nanophotonics}
to appear (Oxford University Press, 2014).

\bibitem{Bordag2011}
M. Bordag, \Review{Eur. Phys. J. C} \Volume{71} 1788 (2011).

\bibitem{Matsubara1955}
T. Matsubara, \Review{Prog. Theor. Phys.} \Volume{14} 351 (1955).

\bibitem{Banishev2012}
A.A. Banishev, C.-C. Chang, G. L. Klimchitskaya, V.M. Mostepanenko and U. Mohideen,
\Review{Phys. Rev. B} \Volume{85} 195422 (2012).

\bibitem{Banishev2013}
A.A. Banishev, G.L. Klimchitskaya,  V.M. Mostepanenko and U. Mohideen,
\Review{Phys. Rev. B} \Volume{88} 155410 (2013).

\bibitem{Behunin2012}
R.O. Behunin, F. Intravaia, D.A.R. Dalvit, P.A. Maia Neto and S. Reynaud,
\Review{Phys. Rev. A} \Volume{85} 012504 (2012).

\bibitem{Broer2012}
W. Broer, G. Palasantzas, J. Knoester and V.B. Svetovoy,
\Review{Phys. Rev. B} \Volume{85} 155410 (2012).

\bibitem{Genet2003}
C. Genet, A. Lambrecht and S. Reynaud,
\Review{Phys. Rev. A} \Volume{67} 043811 (2003).

\bibitem{Lambrecht2006}
A. Lambrecht, P.A. Maia Neto and S. Reynaud,
\Review{New J. Phys.} \Volume{8} 243 (2006).

\bibitem{Schwinger1978}
J. Schwinger, L.L. DeRaad, Jr. and K.A. Milton,
\Review{Ann. Phys.} \Volume{115} 1 (1978).

\bibitem{Intravaia2005}
F. Intravaia and A. Lambrecht,
\Review{Phys. Rev. Lett.} \Volume{94} 110404 (2005).

\bibitem{Intravaia2007}
F. Intravaia, C. Henkel and A. Lambrecht,
\Review{Phys. Rev. A} \Volume{76} 033820 (2007).

\bibitem{Intravaia2009}
F. Intravaia and C. Henkel,
\Review{Phys. Rev. Lett.} \Volume{103} 130405 (2009).

\bibitem{Intravaia2010}
F. Intravaia, S.\AA. Ellingsen and C. Henkel,
\Review{Phys. Rev. A} \Volume{82} 032504 (2010).

\bibitem{Vladimirov1971}
V. S. Vladimirov, \Book{Equations of Mathematical Physics},
(Decker New York, 1971).

\bibitem{Nesterenko2012}
V. Nesterenko and I. Pirozhenko,
\Review{Phys. Rev. A} \Volume{86} 052503 (2012).

\bibitem{Callen1951}
H.B. Callen and T.A. Welton, \Review{Phys. Rev.} \Volume{83} 34
(1951).

\bibitem{Rytov1953}
S.M. Rytov, \Book{Theory of Electric Fluctuations and Thermal
Radiation} (The Academy of Sciences, Moscow, 1953).

\bibitem{Polder1971}
D. Polder and M. Van Hove, \Review{Phys. Rev. B} \Volume{4} 3303
(1971).

\bibitem{Lifshitz1980}
E.M. Lifshitz and L.P. Pitaevskii, \Book{L.D. Landau and E.M.
Lifshitz Course of Theoretical Physics, Statistical Physics Part 2}
\S 77 (1980).

\bibitem{Rytov1989}
S.M. Rytov, Y.A. Kravtsov and V.I. Tatarskii, \Book{Principles of
Statistical Radiophysics} (Springer, 1989).

\bibitem{Antezza2008}
M. Antezza, L.P. Pitaevskii, S. Stringari, and V.B. Svetovoy,
\Review{Phys. Rev. A} \Volume{77} 022901 (2008).

\end{thebibliography}
\end{document}